\documentclass[twocolumn]{aastex7}
\usepackage{hyperref,amsmath, bm, empheq, mathrsfs,  cancel, multirow, xcolor,array,natbib}

\begin{document}

\title{Dynamo Confinement of a Radiatively Spreading Solar Tachocline Revealed by Self-consistent Global Simulations}

\correspondingauthor{Loren I. Matilsky}
\author[0000-0001-9001-6118]{Loren I. Matilsky}\thanks{U.S. National Science Foundation\\ Astronomy and Astrophysics Postdoctoral Fellow}	
\affiliation{Department of Applied Mathematics,
Baskin School of Engineering,
University of California, 
Santa Cruz, CA 96064-1077, USA}
\email{lmatilsk@ucsc.edu}

\author[0000-0002-0963-4881]{Lydia Korre}
\affiliation{Department of Applied Mathematics,
University of Colorado,
Boulder, CO 80309-0526, USA}	
\email{lydia.korre@colorado.edu}

\author[0000-0003-4350-5183]{Nicholas H. Brummell}
\affiliation{Department of Applied Mathematics,
Baskin School of Engineering,
University of California,
Santa Cruz, CA 96064-1077, USA}	
\email{brummell@ucsc.edu}


\newcommand{\testmacro}{\text{macros.tex v. 2025-01-02 6pm}}
\newcommand{\myemail}{loren.matilsky@gmail.com}

\newcommand{\sn}[2]{#1\times10^{#2}}
\newcommand{\sncomp}[2]{{#1}e{#2}}

\newcommand{\cz}{_{\rm{cz}}}
\newcommand{\rz}{_{\rm{rz}}}
\newcommand{\fulll}{_{\rm{full}}}
\newcommand{\dimm}{_{\rm{dim}}}
\newcommand{\rad}{_{\rm{r}}}
\newcommand{\nrad}{_{\rm{nr}}}

\newcommand{\pderiv}[2]{\frac{\partial#1}{\partial#2}}
\newcommand{\matderiv}[1]{\frac{D#1}{Dt}}
\newcommand{\pderivline}[2]{\partial#1/\partial#2}
\newcommand{\parenfrac}[2]{\left(\frac{#1}{#2}\right)}
\newcommand{\brackfrac}[2]{\left[\frac{#1}{#2}\right]}
\newcommand{\bracefrac}[2]{\left\{\frac{#1}{#2}\right\}}

\newcommand{\av}[1]{\left\langle#1\right\rangle}
\newcommand{\avsph}[1]{\left\langle#1\right\rangle_{\rm{sph}}}
\newcommand{\avspht}[1]{\left\langle#1\right\rangle_{ {\rm sph}, t}}
\newcommand{\avt}[1]{\left\langle#1\right\rangle_{t}}
\newcommand{\avtpar}[1]{\left(#1\right)_{t}}
\newcommand{\avphi}[1]{\left\langle#1\right\rangle_{\phi}}
\newcommand{\avphit}[1]{\left\langle#1\right\rangle_{\phi,t}}
\newcommand{\avvol}[1]{\left\langle#1\right\rangle_{\rm{v}}}

\newcommand{\avcz}[1]{\left\langle#1\right\rangle\cz}
\newcommand{\avrz}[1]{\left\langle#1\right\rangle\rz}
\newcommand{\avfull}[1]{\left\langle#1\right\rangle\fulll}
\newcommand{\avczt}[1]{\left\langle#1\right\rangle_{{\rm cz}, t}}
\newcommand{\avrzt}[1]{\left\langle#1\right\rangle{{\rm rz}, t}}
\newcommand{\avfullt}[1]{\left\langle#1\right\rangle{{\rm full}, t}}

\newcommand{\avalt}[1]{\langle#1\rangle}
\newcommand{\avaltsph}{\overline}
\newcommand{\avaltspht}[1]{\left(\overline{#1}\right)_{t}}

\newcommand{\dbprime}{^{\prime\prime}}

\newcommand{\define}{\coloneqq}
\newcommand{\definealt}{\equiv}

\newcommand{\five}{\ \ \ \ \ }
\newcommand{\orr}{\text{or}\five }
\newcommand{\andd}{\text{and}\five }
\newcommand{\where}{\text{where}\five }
\newcommand{\with}{\text{with}\five }


\newcommand{\curl}{\nabla\times}
\newcommand{\Div}{\nabla\cdot}
\newcommand{\lap}{\nabla^2}
\newcommand{\dotgrad}{\cdot\nabla}
\newcommand{\ugrad}{\bm{u}\dotgrad}

\newcommand{\e}{\hat{\bm{e}}}
\newcommand{\erad}{\e_r}
\newcommand{\etheta}{\e_\theta}
\newcommand{\ephi}{\e_\phi}
\newcommand{\elambda}{\e_\lambda}
\newcommand{\ez}{\e_z}
\newcommand{\exi}{\e_\xi}
\newcommand{\eeta}{\e_\eta}
\newcommand{\epol}{\e_{\rm{pol}}}
\newcommand{\emer}{\e_{\rm{mer}}}

\newcommand{\flux}{{\bm{F}}}

\newcommand{\fluxrad}{\flux_{\rm{r}}}
\newcommand{\fluxnrad}{\flux_{\rm{nr}}}
\newcommand{\fluxcond}{\flux_{\rm{c}}}
\newcommand{\fluxenth}{\flux_{\rm{e}}}

\newcommand{\fluxscalarrad}{F_{\rm{r}}}
\newcommand{\fluxscalarnrad}{F_{\rm{nr}}}
\newcommand{\fluxscalarcond}{F_{\rm{c}}}
\newcommand{\fluxscalarenth}{F_{\rm{e}}}

\newcommand{\omzero}{\Omega_0}
\newcommand{\twoomzero}{2\Omega_0}
\newcommand{\omzerovec}{\bm{\Omega}_0}
\newcommand{\omsun}{\Omega_\odot}
\newcommand{\omrz}{\Omega\rz}
\newcommand{\omcz}{\Omega\cz}
\newcommand{\bruntsun}{N_\odot}
\newcommand{\domega}{\Delta\Omega}
\newcommand{\domsixty}{\domega_{\rm 60}}


\newcommand{\ofr}{(r)}
\newcommand{\rprime}{{r^{\prime}}}
\newcommand{\ofrprime}{(\rprime)}

\newcommand{\cv}{c_{\rm{v}}}
\newcommand{\cp}{c_{\rm{p}}}
\newcommand{\cvcap}{C_{\rm{v}}}
\newcommand{\cpcap}{C_{\rm{p}}}
\newcommand{\cs}{c_{\rm s}}
\newcommand{\gasconst}{\mathcal{R}}
\newcommand{\gammaone}{\Gamma_1}

\newcommand{\tot}{_{\rm{tot}}}
\newcommand{\rhotot}{\rho\tot}
\newcommand{\tmptot}{T\tot}
\newcommand{\prstot}{P\tot}
\newcommand{\stot}{S\tot}
\newcommand{\dsdrtot}{\frac{dS\tot}{dr}}
\newcommand{\dsdrtotline}{dS\tot/dr}

\newcommand{\Dentr}{\Delta s}

\newcommand{\rhoover}{\overline{\rho}}
\newcommand{\tmpover}{\overline{T}}
\newcommand{\prsover}{\overline{P}}
\newcommand{\entrover}{\overline{s}}
\newcommand{\inteover}{\overline{u}}
\newcommand{\enthover}{\overline{h}}
\newcommand{\heatover}{\overline{Q}}
\newcommand{\heatradover}{\overline{Q}_{\rm r}}
\newcommand{\coolover}{\overline{C}}
\newcommand{\nsqover}{\overline{N^2}}
\newcommand{\gover}{\overline{g}}
\newcommand{\nuover}{\overline{\nu}}
\newcommand{\kappaover}{\overline{\kappa}}
\newcommand{\etaover}{\overline{\eta}}
\newcommand{\muover}{\overline{\mu}}
\newcommand{\deltaover}{\overline{\delta}}
\newcommand{\cpover}{\overline{\cpcap}}
\newcommand{\cvover}{\overline{\cvcap}}
\newcommand{\cssqover}{\overline{\cs^2}}
\newcommand{\Dentrover}{\Delta\entrover}
\newcommand{\Dentroverf}{\Delta\entrover_{\rm f}}

\newcommand{\fluxradover}{\overline{\flux}_{\rm{r}}}
\newcommand{\fluxscalarradover}{\overline{F}_{\rm{r}}}
\newcommand{\fluxnradover}{\overline{\flux}_{\rm{nr}}}
\newcommand{\fluxscalarnradover}{\overline{F}_{\rm{nr}}}
\newcommand{\kradover}{\overline{\kappa}_{\rm rad}}

\newcommand{\rhoa}{\rho_a}
\newcommand{\rhocz}{\rho\cz}
\newcommand{\rhorz}{\rho\rz}
\newcommand{\tmpa}{T_a}
\newcommand{\tmpcz}{T\cz}
\newcommand{\tmprz}{T\rz}
\newcommand{\prsa}{p_a}
\newcommand{\prscz}{p\cz}
\newcommand{\prsrz}{p\rz}
\newcommand{\entra}{s_a}
\newcommand{\intea}{u_a}
\newcommand{\entha}{h_a}
\newcommand{\heata}{Q_a}
\newcommand{\heatrada}{\overline{Q}_{{\rm r}a}}
\newcommand{\coola}{C_a}
\newcommand{\nsqa}{N^2_a}
\newcommand{\nsqrz}{N^2\rz}
\newcommand{\bruntrz}{N\rz}
\newcommand{\grava}{g_a}
\newcommand{\gravcz}{g\cz}
\newcommand{\gravrz}{g\rz}
\newcommand{\nua}{\nu_a}
\newcommand{\nut}{\nu_{\rm t}}
\newcommand{\nurz}{\nu\rz}
\newcommand{\nucz}{\nu\cz}
\newcommand{\kappaa}{\kappa_a}
\newcommand{\kappat}{\kappa_{\rm t}}
\newcommand{\kapparz}{\kappa\rz}
\newcommand{\kappacz}{\kappa\cz}

\newcommand{\etaa}{\eta_a}
\newcommand{\etarz}{\eta\rz}
\newcommand{\etacz}{\eta\cz}
\newcommand{\mua}{\mu_a}
\newcommand{\deltaa}{\delta_a}
\newcommand{\cpa}{(\cpcap)_a}
\newcommand{\cva}{(\cvcap)_a}
\newcommand{\cssqa}{(\cs^2)_a}
\newcommand{\fluxrada}{F_{{\rm r}a}}
\newcommand{\fluxnrada}{F_{{\rm{nr}}a}}
\newcommand{\fluxnradcz}{F_{{\rm{nr,cz}}}}
\newcommand{\krada}{\kappa_{{\rm rad}a}}
\newcommand{\sigmaa}{\sigma_a}

\newcommand{\dlnrhoover}{\frac{d\ln\rhoover}{dr}}
\newcommand{\dlntmpover}{\frac{d\ln\tmpover}{dr}}
\newcommand{\dlnprsover}{\frac{d\ln\prsover}{dr}}
\newcommand{\dsdrover}{\frac{d\entrover}{dr}}
\newcommand{\dsdroverline}{d\entrover/dr}
\newcommand{\dlnrhooverline}{d\ln\rhoover/dr}
\newcommand{\dlntmpoverline}{d\ln\tmpover/dr}
\newcommand{\dlnprsoverline}{d\ln\prsover/dr}

\newcommand{\rhotilde}{\tilde{\rho}}
\newcommand{\tmptilde}{\tilde{T}}
\newcommand{\prstilde}{\tilde{p}}
\newcommand{\entrtilde}{\tilde{s}}
\newcommand{\intetilde}{\tilde{u}}
\newcommand{\enthtilde}{\tilde{h}}
\newcommand{\heattilde}{\tilde{Q}}
\newcommand{\heatradtilde}{\tilde{Q}_{\rm r}}
\newcommand{\dentr}{\delta_s}
\newcommand{\dheat}{\delta_Q}
\newcommand{\drad}{\delta_{\rm r}}
\newcommand{\rheat}{r_Q}
\newcommand{\rentr}{r_s}
\newcommand{\lumtilde}{\tilde{L}}
\newcommand{\heatra}{\tilde{Q}_{\texttt{Ra}}}
\newcommand{\cooltilde}{\tilde{C}}
\newcommand{\brunttilde}{\widetilde{N}}
\newcommand{\brunttildesq}{\widetilde{N}^2}
\newcommand{\nsqtilde}{\widetilde{N^2}}
\newcommand{\gtilde}{\tilde{g}}
\newcommand{\nutilde}{\tilde{\nu}}
\newcommand{\kappatilde}{\tilde{\kappa}}
\newcommand{\etatilde}{\tilde{\eta}}
\newcommand{\mutilde}{\tilde{\mu}}
\newcommand{\deltatilde}{\tilde{\delta}}
\newcommand{\cptilde}{\tilde{\cpcap}}
\newcommand{\cvtilde}{\tilde{\cvcap}}
\newcommand{\cssqtilde}{\tilde{\cs^2}}
\newcommand{\Dentrtilde}{\Delta\entrtilde}
\newcommand{\Dentrtildef}{(\Delta\entrtilde)_{\rm f}}

\newcommand{\fluxtilde}{\tilde{\flux}}
\newcommand{\fluxscalartilde}{\tilde{F}}
\newcommand{\fluxradtilde}{\tilde{\flux}_{\rm{r}}}
\newcommand{\fluxscalarradtilde}{\tilde{F}_{\rm{r}}}
\newcommand{\fluxnradtilde}{\tilde{\flux}_{\rm{nr}}}
\newcommand{\fluxscalarnradtilde}{\tilde{F}_{\rm{nr}}}
\newcommand{\fluxcondtilde}{\tilde{\flux}_{\rm{c}}}
\newcommand{\fluxscalarcondtilde}{\tilde{F}_{\rm{c}}}

\newcommand{\dlnrhotilde}{\frac{d\ln\rhotilde}{dr}}
\newcommand{\dlntmptilde}{\frac{d\ln\tmptilde}{dr}}
\newcommand{\dlnprstilde}{\frac{d\ln\prstilde}{dr}}
\newcommand{\dsdrtilde}{\frac{d\entrtilde}{dr}}
\newcommand{\dlnrhotildeline}{d\ln\rhotilde/dr}
\newcommand{\dlntmptildeline}{d\ln\tmptilde/dr}
\newcommand{\dlnprstildeline}{d\ln\prstilde/dr}
\newcommand{\dsdrtildeline}{d\entrtilde/dr}

\newcommand{\grav}{g}
\newcommand{\vecg}{\bm{g}}
\newcommand{\geff}{g_{\rm{eff}}}
\newcommand{\vecgeff}{\bm{g}_{\rm{eff}}}
\newcommand{\heat}{Q}
\newcommand{\buoyfreq}{N}
\newcommand{\brunt}{N}
\newcommand{\nsq}{N^2}
\newcommand{\hrho}{H_\rho}
\newcommand{\hprs}{H_{\rm{p}}}
\newcommand{\hrhotilde}{\widetilde{H_\rho}}
\newcommand{\hprstilde}{\widetilde{H_{\rm p}}}

\newcommand{\gradrad}{\nabla_{\rm r}}
\newcommand{\gradad}{\nabla_{\rm ad}}

\newcommand{\rhoprime}{{\rho^\prime}}
\newcommand{\tmpprime}{{T^\prime}}
\newcommand{\prsprime}{{p^\prime}}
\newcommand{\entrprime}{{s^\prime}}
\newcommand{\inteprime}{{u^\prime}}
\newcommand{\heatprime}{{Q^\prime}}
\newcommand{\enthprime}{{h^\prime}}
\newcommand{\fradprime}{\bm{F}^\prime_{\rm rad}}
\newcommand{\kradprime}{\kappa^\prime_{\rm rad}}
\newcommand{\fcondprime}{\bm{F}^\prime_{\rm cond}}

\newcommand{\rhohat}{\hat{\rho}}
\newcommand{\tmphat}{\hat{T}}
\newcommand{\prshat}{\hat{p}}
\newcommand{\entrhat}{\hat{s}}
\newcommand{\intehat}{\hat{u}}
\newcommand{\enthhat}{\hat{h}}

\newcommand{\rhoone}{\rho_1}
\newcommand{\tmpone}{T_1}
\newcommand{\prsone}{p_1}
\newcommand{\entrone}{s_1}
\newcommand{\inteone}{u_1}
\newcommand{\enthone}{h_1}

\newcommand{\pomega}{\varpi}

\newcommand{\vecu}{\bm{u}}
\newcommand{\vecb}{\bm{B}}
\newcommand{\vecom}{\bm{\omega}}
\newcommand{\vecj}{\bm{\mathcal{J}}}

\newcommand{\upol}{\vecu_{\rm{pol}}}
\newcommand{\bpol}{\vecb_{\rm{pol}}}
\newcommand{\umer}{\vecu_{\rm{m}}}
\newcommand{\bmer}{\vecb_{\rm{m}}}

\newcommand{\urad}{u_r}
\newcommand{\utheta}{u_\theta}
\newcommand{\uphi}{u_\phi}
\newcommand{\ulambda}{u_\lambda}
\newcommand{\uz}{u_z}

\newcommand{\rhoumer}{\av{\rhotilde\umer}}
\newcommand{\rhourad}{\av{\rhotilde\urad}}
\newcommand{\rhoutheta}{\av{\rhotilde\utheta}}
\newcommand{\rhoulambda}{\av{\rhotilde\ulambda}}
\newcommand{\rhouz}{\av{\rhotilde\uz}}

\newcommand{\omrad}{\omega_r}
\newcommand{\omtheta}{\omega_\theta}
\newcommand{\omphi}{\omega_\phi}
\newcommand{\omlambda}{\omega_\lambda}
\newcommand{\omz}{\omega_z}

\newcommand{\brad}{B_r}
\newcommand{\btheta}{B_\theta}
\newcommand{\bphi}{B_\phi}
\newcommand{\blambda}{B_\lambda}
\newcommand{\bz}{B_z}

\newcommand{\jrad}{\mathcal{J}_r}
\newcommand{\jtheta}{\mathcal{J}_\theta}
\newcommand{\jphi}{\mathcal{J}_\phi}
\newcommand{\jlambda}{\mathcal{J}_\lambda}
\newcommand{\jz}{\mathcal{J}_z}

\newcommand{\vecuprime}{\bm{u}^\prime}
\newcommand{\vecbprime}{\bm{B}^\prime}
\newcommand{\vecomprime}{\bm{\omega}^\prime}
\newcommand{\vecjprime}{\bm{\mathcal{J}}^\prime}

\newcommand{\upolprime}{\vecu_{\rm{pol}}^\prime}
\newcommand{\bpolprime}{\vecb_{\rm{pol}}^\prime}
\newcommand{\umerprime}{\vecu_{\rm{m}}^\prime}
\newcommand{\bmerprime}{\vecb_{\rm{m}}^\prime}

\newcommand{\uradprime}{u_r^\prime}
\newcommand{\uthetaprime}{u_\theta^\prime}
\newcommand{\uphiprime}{u_\phi^\prime}
\newcommand{\ulambdaprime}{u_\lambda^\prime}
\newcommand{\uzprime}{u_z^\prime}

\newcommand{\avurad}{\av{u_r}}
\newcommand{\avutheta}{\av{u_\theta}}
\newcommand{\avuphi}{\av{u_\phi}}
\newcommand{\avulambda}{\av{u_\lambda}}
\newcommand{\avuz}{\av{u_z}}
\newcommand{\aventrhat}{\av{\entrhat}}

\newcommand{\omradprime}{\omega_r^\prime}
\newcommand{\omthetaprime}{\omega_\theta^\prime}
\newcommand{\omphiprime}{\omega_\phi^\prime}
\newcommand{\omlambdaprime}{\omega_\lambda^\prime}
\newcommand{\omzprime}{\omega_z^\prime}

\newcommand{\bradprime}{B_r^\prime}
\newcommand{\bthetaprime}{B_\theta^\prime}
\newcommand{\bphiprime}{B_\phi^\prime}
\newcommand{\blambdaprime}{B_\lambda^\prime}
\newcommand{\bzprime}{B_z^\prime}

\newcommand{\jradprime}{\mathcal{J}_r^\prime}
\newcommand{\jthetaprime}{\mathcal{J}_\theta^\prime}
\newcommand{\jphiprime}{\mathcal{J}_\phi^\prime}
\newcommand{\jlambdaprime}{\mathcal{J}_\lambda^\prime}
\newcommand{\jzprime}{\mathcal{J}_z^\prime}

\newcommand{\cost}{\cos\theta}
\newcommand{\sint}{\sin\theta}
\newcommand{\cott}{\cot\theta}
\newcommand{\rsint}{r\sint}
\newcommand{\orsint}{\frac{1}{\rsint}}
\newcommand{\orsintline}{(1/\rsint)}
\newcommand{\rt}{r\theta}

\newcommand{\amom}{\mathcal{L}}

\newcommand{\minn}{_{\rm{min}}}
\newcommand{\maxx}{_{\rm{max}}}
\newcommand{\inn}{_{\rm{in}}}
\newcommand{\out}{_{\rm{out}}}
\newcommand{\bott}{_{\rm{bot}}}
\newcommand{\midd}{_{\rm{mid}}}
\newcommand{\topp}{_{\rm{top}}}
\newcommand{\bcz}{_{\rm{bcz}}}
\newcommand{\ov}{_{\rm{ov}}}
\newcommand{\rms}{_{\rm{rms}}}
\newcommand{\const}{_{\rm{const}}}

\newcommand{\nr}{N_r}
\newcommand{\nt}{N_\theta}
\newcommand{\np}{N_\phi}
\newcommand{\nmax}{{n_{\rm{max}}}}
\newcommand{\lmax}{{\ell_{\rm{max}}}}

\newcommand{\lsun}{L_\odot}

\newcommand{\msun}{M_\odot}
\newcommand{\rstar}{R_*}
\newcommand{\lstar}{L_*}
\newcommand{\mstar}{M_*}

\newcommand{\rearth}{R_\oplus}
\newcommand{\omearth}{\Omega_\oplus}
\newcommand{\mearth}{M_\oplus}

\newcommand{\taurs}{\tau_{\rm{rs}}}
\newcommand{\taumc}{\tau_{\rm{mc}}}
\newcommand{\tauv}{\tau_{\rm{v}}}
\newcommand{\taurad}{\tau_{\rm{rad}}}
\newcommand{\taums}{\tau_{\rm{ms}}}
\newcommand{\taumm}{\tau_{\rm{mm}}}
\newcommand{\taumag}{\tau_{\rm{mag}}}

\newcommand{\aflux}{\bm{\mathcal{F}}}
\newcommand{\afluxrs}{\aflux^{\rm{rs}}}
\newcommand{\afluxmc}{\aflux^{\rm{mc}}}
\newcommand{\afluxv}{\aflux^{\rm{v}}}
\newcommand{\afluxms}{\aflux^{\rm{ms}}}
\newcommand{\afluxmm}{\aflux^{\rm{mm}}}
\newcommand{\afluxmag}{\aflux^{\rm{mag}}}

\newcommand{\taunu}{\tau_{\nu}}
\newcommand{\taukappa}{\tau_{\kappa}}
\newcommand{\taueta}{\tau_{\eta}}
\newcommand{\tauff}{\tau_{\rm ff}}
\newcommand{\tauomega}{\tau_\Omega}
\newcommand{\taun}{\tau_N}
\newcommand{\taues}{\tau_{\rm ES}}

\newcommand{\pes}{{P_{\rm{ES}}}}
\newcommand{\pburrow}{{P_{\rm{b}}}}
\newcommand{\pessun}{{P_{ {\rm ES}, \odot}}}
\newcommand{\pnu}{{P_{\nu}}}
\newcommand{\pkappa}{{P_{\kappa}}}
\newcommand{\peta}{{P_{\eta}}}
\newcommand{\prot}{{P_{\rm{rot}}}}
\newcommand{\pom}{{P_{\Omega}}}
\newcommand{\pff}{{P_{\rm ff}}}
\newcommand{\pequil}{{P_{\rm{eq}}}}
\newcommand{\pcyc}{{P_{\rm{cyc}}}}
\newcommand{\pcycm}{{P_{{\rm cyc}, m}}}

\newcommand{\pesnd}{{\hat{P}_{\rm{ES}}}}
\newcommand{\pburrownd}{{\hat{P}_{\rm{b}}}}
\newcommand{\pessunnd}{{\hat{P}_{ {\rm ES}, \odot}}}
\newcommand{\pnund}{{\hat{P}_{\nu}}}
\newcommand{\pkappand}{{\hat{P}_{\kappa}}}
\newcommand{\petand}{{\hat{P}_{\eta}}}
\newcommand{\protnd}{{\hat{P}_{\rm{rot}}}}
\newcommand{\pomnd}{{\hat{P}_{\Omega}}}
\newcommand{\pffnd}{{\hat{P}_{\rm ff}}}
\newcommand{\pequilnd}{{\hat{P}_{\rm{eq}}}}
\newcommand{\pcycnd}{{\hat{P}_{\rm{cyc}}}}
\newcommand{\pcycmnd}{{\hat{P}_{{\rm cyc}, m}}}

\newcommand{\tes}{{t_{\rm{es}}}}
\newcommand{\test}{{t_{\rm{es,t}}}}
\newcommand{\tcirc}{{t_{\rm{circ}}}}
\newcommand{\tburrow}{{t_{\rm{b}}}}
\newcommand{\tessun}{{t_{ {\rm es}, \odot}}}
\newcommand{\tnu}{{t_{\nu}}}
\newcommand{\tnut}{{t_{{\nu}\rm,t}}}
\newcommand{\tkappa}{{t_{\kappa}}}
\newcommand{\teta}{{t_{\eta}}}
\newcommand{\trot}{{t_{\rm{rot}}}}
\newcommand{\tomega}{{t_{\Omega}}}
\newcommand{\tbrunt}{t_N}
\newcommand{\tvs}{{t_{\rm vs}}}
\newcommand{\trs}{{t_{\rm rs}}}

\newcommand{\tff}{{t_{\rm ff}}}
\newcommand{\tequil}{{t_{\rm{eq}}}}
\newcommand{\tcyc}{{t_{\rm{cyc}}}}
\newcommand{\trun}{{t_{\rm{run}}}}
\newcommand{\tmax}{{t_{\rm{max}}}}
\newcommand{\tcycm}{{t_{{\rm cyc}, m}}}
\newcommand{\tsun}{t_\odot}

\newcommand{\tesdim}{{t_{\rm{es}}^*}}
\newcommand{\testdim}{{t_{\rm{es,t}}^*}}
\newcommand{\tcircdim}{{t_{\rm{circ}}^*}}
\newcommand{\tesshdim}{{t_{\rm{es,sh}}^*}}
\newcommand{\tburrowdim}{{t_{\rm{b}}^*}}
\newcommand{\tnudim}{{t_{\nu}^*}}
\newcommand{\tnutdim}{{t_{{\nu}\rm,t}^*}}
\newcommand{\tkappadim}{{t_{\kappa}^*}}
\newcommand{\tetadim}{{t_{\eta}^*}}
\newcommand{\trotdim}{{t_{\rm{rot}}^*}}
\newcommand{\tomegadim}{{t_{\Omega}^*}}
\newcommand{\tbruntdim}{{t_N^*}}

\newcommand{\tffdim}{{t_{\rm ff}^*}}
\newcommand{\tequildim}{{t_{\rm{eq}}^*}}
\newcommand{\tcycdim}{{t_{\rm{cyc}}^*}}
\newcommand{\trundim}{{t_{\rm{run}}^*}}
\newcommand{\tmaxdim}{{t_{\rm{max}}^*}}
\newcommand{\tsundim}{t_{\odot}^*}
\newcommand{\tvsdim}{t_{\rm vs}^*}
\newcommand{\trsdim}{t_{\rm rs}^*}

\newcommand{\omcyc}{{\omega_{\rm{cyc}}}}
\newcommand{\omcycm}{{\omega_{{\rm{cyc}}, m}}}

\newcommand{\ra}{{\rm{Ra}}}
\newcommand{\ratwo}{{\rm{Ra_2}}}
\newcommand{\sigmaeff}{{\sigma_{\rm eff}}}
\newcommand{\sigmaest}{{\sigma_{\rm est}}}
\newcommand{\sigmaeffloc}{{\sigma_{\rm eff,loc}}}
\newcommand{\sigmazero}{{\sigma_0}}
\newcommand{\sigmazerosq}{{\sigma_0^2}}
\newcommand{\sigmashsq}{{\sigma^2_{\rm sh}}}
\newcommand{\sigmaloc}{{\sigma_{\rm loc}}}
\newcommand{\sigmadyn}{{\sigma_{\rm dyn}}}
\newcommand{\raf}{\ra_{\rm{f}}}
\newcommand{\rafloc}{\ra_{\rm{f,loc}}}
\newcommand{\ramod}{\ra^*}
\newcommand{\rafmod}{\raf^*}
\newcommand{\pr}{{\rm{Pr}}}
\newcommand{\prm}{{\rm{Pr_m}}}
\newcommand{\ek}{{\rm{Ek}}}
\newcommand{\ekt}{{\rm{Ek_T}}}
\newcommand{\ta}{{\rm{Ta}}}
\newcommand{\roc}{{\rm{Ro_c}}}
\newcommand{\rocsq}{{\rm{Ro_c^2}}}
\newcommand{\bu}{{\rm{Bu}}}
\newcommand{\bumod}{{\rm{Bu^*}}}
\newcommand{\buvisc}{{\rm{Bu_{visc}}}}
\newcommand{\burot}{{\rm{Bu_{rot}}}}
\newcommand{\fr}{{\rm Fr}}

\newcommand{\Nrho}{N_\rho}
\newcommand{\nrho}{n_\rho}
\newcommand{\di}{{\rm{Di}}}


\newcommand{\ro}{{\rm{Ro}}}
\newcommand{\lo}{{\rm{Lo}}}

\newcommand{\re}{{\rm{Re}}}
\newcommand{\rem}{{\rm{Re_m}}}

\newcommand{\gram}{{\rm{g}}}
\newcommand{\cm}{{\rm{cm}}}
\newcommand{\second}{{\rm{s}}}
\newcommand{\radsecond}{{\rm rad\ s^{-1}}}

\newcommand{\yr}{{\rm{yr}}}
\newcommand{\gauss}{{\rm{G}}}
\newcommand{\kelv}{{\rm{K}}}
\newcommand{\unitentr}{{\rm{erg\ g^{-1}\ K^{-1}}}}
\newcommand{\unitdsdr}{{\rm{erg\ g^{-1}\ K^{-1}\ cm^{-1}}}}
\newcommand{\uniten}{\rm{erg}\ \cm^{-3}}
\newcommand{\unitprs}{\rm{dyn}\ \cm^{-2}}
\newcommand{\unitrho}{\gram\ \cm^{-3}}
\newcommand{\stoke}{\rm{cm^2\ s^{-1}}}

\newcommand{\meanb}{\overline{\bm{B}}}
\newcommand{\flucb}{\bm{B}^\prime}
\newcommand{\totb}{\bm{B}}

\newcommand{\meanv}{\overline{\bm{v}}}
\newcommand{\flucv}{\bm{v}^\prime}
\newcommand{\totv}{\bm{v}}

\newcommand{\emf}{\bm{\mathcal{E}}}
\newcommand{\meanemf}{\overline{\bm{\mathcal{E}}}}
\newcommand{\meanbpol}{\overline{\bm{B}_{\rm{pol}}}}


\newcommand{\rin}{r_{\rm i}}
\newcommand{\rout}{r_{\rm o}}
\newcommand{\rbot}{r_{\rm b}}
\newcommand{\rtop}{r_{\rm t}}

\newcommand{\rindim}{r_{\rm i}^*}
\newcommand{\routdim}{r_{\rm o}^*}
\newcommand{\rbotdim}{r_{\rm b}^*}
\newcommand{\rtopdim}{r_{\rm t}^*}

\newcommand{\rbcz}{r_{\rm bcz}}
\newcommand{\rtcz}{r_{\rm tcz}}
\newcommand{\rbrz}{r_{\rm brz}}
\newcommand{\rtrz}{r_{\rm trz}}
\newcommand{\rc}{r_{\rm c}}
\newcommand{\rnrhothree}{r_{3}}
\newcommand{\rtach}{r_{\rm t}}
\newcommand{\dtach}{\Delta_{\rm t}}
\newcommand{\dtachsun}{\Delta_\odot}

\newcommand{\rbczdim}{{r_{\rm bcz}^*}}
\newcommand{\rtczdim}{{r_{\rm tcz}^*}}
\newcommand{\rbrzdim}{{r_{\rm brz}^*}}
\newcommand{\rtrzdim}{{r_{\rm trz}^*}}
\newcommand{\rcdim}{{r_{\rm c}^*}}
\newcommand{\rnrhothreedim}{{r_{3}^*}}
\newcommand{\rtachdim}{{r_{\rm t}^*}}
\newcommand{\dtachdim}{{\Delta_{\rm t}^*}}

\newcommand{\rsun}{R_\odot}
\newcommand{\rbczsun}{R_{\rm bcz}}
\newcommand{\rnrhothreesun}{R_{\rm 3}}
\newcommand{\rtachsun}{R_{\rm t}}

\newcommand{\rayleigh}{\texttt{Rayleigh}}
\newcommand{\rayleigha}{\texttt{Rayleigh 0.9.1}}
\newcommand{\rayleighb}{\texttt{Rayleigh 1.0.1}}

\newcommand{\eulag}{\texttt{EULAG}}
\newcommand{\eulagmhd}{\texttt{EULAG-MHD}}
\newcommand{\ash}{\texttt{ASH}}
\newcommand{\rsst}{\texttt{RSST}}
\newcommand{\rtdt}{\texttt{R2D2}}
\newcommand{\pencil}{\texttt{Pencil}}

\defcitealias{Clark1973}{C73}
\defcitealias{Spiegel1992}{SZ92}
\defcitealias{Gough1998}{GM98}
\defcitealias{Matilsky2022}{M22}
\defcitealias{Matilsky2024}{M24}
\defcitealias{Korre2024}{KF24}
\defcitealias{ForgcsDajka2001}{FDP01}

\renewcommand{\dimm}{^*}
\newcommand{\dimsq}{^{*2}}

\newcommand{\betarz}{\beta\rz}

\newcommand{\lm}[1]{{\color{red}#1}}
\newcommand{\lk}[1]{{\color{blue}#1}}
\newcommand{\nb}[1]{{\color{orange}#1}}

\newcommand\lmout{\bgroup\markoverwith{\textcolor{red}{\rule[0.5ex]{2pt}{1pt}}}\ULon}
\newcommand\lkout{\bgroup\markoverwith{\textcolor{blue}{\rule[0.5ex]{2pt}{1pt}}}\ULon}
\newcommand\nbout{\bgroup\markoverwith{\textcolor{orange}{\rule[0.5ex]{2pt}{1pt}}}\ULon}

\newcommand{\mwsuper}{^{m\omega}}
\newcommand{\bpolmw}{\bpol\mwsuper}
\newcommand{\bthetamw}{\btheta\mwsuper}
\newcommand{\deltamw}{\delta\mwsuper}
\newcommand{\com}[1]{{\color{purple}#1}}

\begin{abstract}

The helioseismically observed solar tachocline is a thin internal boundary layer of shear that separates the rigidly-rotating solar radiative zone from the differentially-rotating convective zone and is believed to play a central role in the 22-year solar dynamo cycle. The observed thinness of the tachocline has long been a mystery, given the expected tendency of such shear to undergo radiative spreading. Radiative spreading is the process by which the meridional circulation and angular velocity burrow into a stably-stratified fluid owing to the mitigating effect of radiative thermal diffusion. A confinement mechanism is thus required to keep the tachocline so thin. In previous work using global dynamo simulations, we achieved a statistically-stationary confined tachocline where the confinement mechanism was derived from the Maxwell stress arising from a dynamo-generated nonaxisymmetric poloidal magnetic field. However, the parameters chosen meant that the tachocline was confined against viscous spread instead of radiative spread. Here, we show that the previously identified dynamo confinement mechanism still succeeds in a simulation that lies in the more solar-like radiative spreading regime. In particular, a nonaxisymmetric, quasicyclic dynamo develops in the convective zone and overshoot layer, penetrates into the radiative zone via a novel type of skin effect, and creates a Maxwell stress that confines the tachocline over many magnetic cycles. To the best of our knowledge, this is the first fully self-consistent rendering of a confined tachocline in a global numerical simulation in the parameter regime appropriate to the Sun.

\end{abstract}



\section{The Solar Dynamo and Tachocline} 

Two primary observations of the Sun indicate large-scale order within the solar interior.  The first is the 22-year dynamo cycle associated with emerging sunspots (e.g., \citealt{Hathaway2015}). The second is the helioseismically inferred internal rotation profile (e.g., \citealt{Howe2009,Basu2016}), whose large-scale shear is thought to be a key driver of the dynamo. The region of strongest shear is the solar tachocline at the base of the convective zone (CZ), where the strong latitudinal differential rotation of the CZ rapidly transitions to the rigid rotation of the radiative zone (RZ) beneath. Understanding this large-scale structure in the magnetic field and rotation profile is one of the primary goals of solar physics. 

The dynamics of the tachocline are particularly important, as it has long been invoked as the ``seat" of the solar dynamo, i.e., the primary location where large-scale poloidal magnetic field is sheared into the toroidal field. Production of toroidal field constitutes one half of the dynamo loop and the tachocline is thus thought to be a possible deep source for sunspots and related magnetic activity (e.g., \citealt{Parker1993,Charbonneau1997}.) The tachocline's existence as a thin, stationary layer is, however, particularly puzzling. \citet{Spiegel1992} (hereafter \citetalias{Spiegel1992}) elucidated a process called ``radiative spreading", in which the baroclinicity associated with the CZ's thermal-wind balance (e.g., \citealt{Matilsky2023}) spreads inward. In radiative spread, the thermal diffusion acts to mitigate the very strong stable stratification of the RZ, which would otherwise keep the shear from burrowing beyond a very narrow rotational shear layer (e.g., \citealt{Holton1965}). On evolutionary timescales, the baroclinicity, induced meridional circulation, and ultimately the differential rotation, are all expected to burrow very deeply into the RZ (see also \citealt{Clark1973,Haynes1991}). An originally thin solar tachocline should have thickened considerably, spreading inward to a radius of about $0.3\rsun$ by the current age of the Sun (see also \citealt{Elliott1997}). Since a deep tachocline is not what we observe, there must be a torque in the RZ which opposes the radiative spread of shear and confines the tachocline to a thin layer. This torque has been hypothesized to arise from hydrodynamic (HD) shear instabilities (\citetalias{Spiegel1992}) or from a primordial magnetic field \citep{Gough1998}, but its precise origins are still not well understood. 

\citet{ForgcsDajka2001} (hereafter \citetalias{ForgcsDajka2001}) proposed that the cycling solar dynamo could also produce such a confining torque. In an idealized two-dimensional model excluding convection and dynamo action, they showed that an axisymmetric poloidal magnetic field undergoing regular polarity reversals could penetrate by at least a skin-depth into the RZ. For turbulently enhanced values of the magnetic diffusivity, this skin-depth could be sufficiently thick that a magnetic confinement mechanism would be able to reproduce the observed thickness of the solar tachocline. 

\citet{Matilsky2022} (hereafter \citetalias{Matilsky2022}) generalized \citetalias{ForgcsDajka2001}'s idealized model significantly using global simulations. \citetalias{Matilsky2022} demonstrated what we now call a dynamo confinement scenario, i.e., the general process by which the magnetic torque from a self-excited dynamo confines a tachocline. Using three-dimensional, fully-nonlinear simulations of convection and dynamo action, \citetalias{Matilsky2022} showed that a convectively-driven dynamo, whose poloidal field was primarily nonaxisymmetric and reversed polarity only quasiregularly, penetrated into the top of the RZ and produced Maxwell stresses that self-consistently confined a tachocline. However, the parameters of these simulations were such that the shear spread into the RZ viscously, instead of radiatively. The central goal of this Letter is to report on new simulations in which dynamo confinement works against radiative spreading.

Despite being in the incorrect parameter regime for the Sun, the mechanism found in \citetalias{Matilsky2022} is a far more flexible incarnation of the skin-depth idea of \citetalias{ForgcsDajka2001}.  A subsequent paper \citet{Matilsky2024} (hereafter \citetalias{Matilsky2024}) showed that because the polarity reversals were only quasiregular and hence occurred with a wide range of frequencies, the depth to which the dynamo field could penetrate was determined by a correspondingly wide range of skin-depths. Furthermore, because of the nonaxisymmetry, the frequencies determining the skin depths were Doppler-shifted by the rigid rotation rate of the RZ and therefore any nonaxisymmetric field almost corotating with the RZ could conceivably penetrate to considerable depth. 

Here in this Letter, we show that the dynamo confinement scenario identified in \citetalias{Matilsky2022} and \citetalias{Matilsky2024} operates successfully in the more solar-appropriate radiative spreading regime. Note that here, we present simply an example of our findings demonstrating the major result, and leave a detailed description of our full simulation suite (which includes a broader survey of the parameter space and a more detailed description of the dynamics) to an upcoming longer article. 

\section{A Statistically-Steady Tachocline Confined against Radiative Spread} 
We use the {\rayleigh} code \citep{Matsui2016,Featherstone2016a} to evolve the anelastic equations of resistive magnetohydrodynamics (MHD; e.g., \citealt{Clune1999,Brun2004}) in a spherical-shell CZ--RZ system, rotating with background angular velocity $\omzero$ about the $z$-axis ($z=r\cos\theta$, where $r$ is the radial coordinate and $\theta$ is the colatitude). The shell extends from inner boundary $\rin=0.491\rsun$ to outer boundary $\rout=0.947\rsun$ (where $\rsun$ is the solar radius), thus capturing the lowermost $\sim$3 density scale heights of the solar CZ and the uppermost $\sim$2 scale heights of the RZ according to the standard solar structure Model S \citep{ChristensenDalsgaard1996}. The internal CZ--RZ boundary, across which overshoot occurs, is located at the shell's midpoint radius $\rc=0.720\rsun$. 

The new simulations described here are designed to assess the robustness of the dynamo confinement scenario of \citetalias{Matilsky2022} and \citetalias{Matilsky2024} under the original conditions for radiative spread described by \citetalias{Spiegel1992}. In \citetalias{Spiegel1992}, the tachocline was infinitely thin initially. The RZ rotated rigidly and was assumed to be completely decoupled from the CZ, save for a latitudinal differential rotation imposed as an upper boundary condition. \citetalias{Spiegel1992} analytically solved for the evolution of the tachocline and showed that radiative spread would occur on the Eddington-Sweet timescale $\tes\define\bu\, \rc^2/\kappa$ \citep{Eddington1925,Sweet1950}. Here, $\kappa$ is the radiative thermal diffusivity and the buoyancy number
\begin{align}\label{eq:bu}
\bu\define\frac{N^2}{4\Omega^2}
\end{align}
(where $\brunt$ is the buoyancy frequency and $\Omega$ is the rotation rate) measures the resistance of the RZ to the rotationally-induced meridional circulation (e.g., \citealt{Holton1965, Barcilon1967}). Viscous spread, on the other hand, occurs on the viscous diffusion timescale $\tnu\define\rc^2/\nu$, where $\nu$ is the viscosity. The Prandtl number
\begin{align}\label{eq:pr}
\pr\define\frac{\nu}{\kappa},
\end{align}
which relates the radiative and viscous timescales, thus plays a role in addition to $\bu$. Ultimately, whether the tachocline spreads viscously or radiatively is determined by a dimensionless quantity (e.g., \citealt{Clark1973}) now called the $\sigma$ parameter (e.g., \citealt{AcevedoArreguin2013}), which characterizes the ratio of the two relevant timescales:
\begin{align}\label{eq:sigma}
\sigma\define \sqrt{\frac{\tes}{\tnu}} = \frac{N}{2\Omega}\sqrt{\frac{\nu}{\kappa}} = \sqrt{\bu\pr}. 
\end{align}
In general, if $\sigma>\sigma_c$, where $\sigma_c$ is an order-unity critical value, viscous spread dominates, whereas if $\sigma<\sigma_c$, radiative spread dominates. The value of $\sigma_c$ depends on various geometric properties of the system and here seems to be around $5$--$10$, although we leave the slightly involved details of this estimate to our followup work.

Using the solar tachocline values from \citealt{Gough2007} for the dimensional quantities in Equations \eqref{eq:bu}--\eqref{eq:sigma} yields $\bu\approx\sn{2}{4}$, $\pr\approx\sn{2}{-6}$, and $\sigma\approx0.2$. The solar RZ is thus firmly in the radiative spreading regime, despite being very stably-stratified (high \bu), by virtue of having extremely low $\pr$. Simulations would, of course, ideally be performed in a similar parameter regime, but the extreme turbulence at low $\pr$ makes the required computational resolution too expensive. Most prior work in global CZ--RZ spherical-shell simulations has been focused on strong stratification ($\bu\gg1$) and $\pr=O(1)$, and thus $\sigma\gg\sigma_c$, leading to negligible radiative spreading. The exception is \citet{Korre2024}, who achieved radiative spreading in weakly stratified simulations with $\pr=1$, $\bu<1$, and thus $\sigma<1$. Here, we take advantage of the fact that $\sigma_c \sim 5$--$10$ to perform more solar-like simulations with $\bu>1$, $\pr<1$, and $1<\sigma<\sigma_c$. We verify that radiative spreading indeed dominates in these simulations after the fact.

To capture the essence of a solar radiative spreading process, we begin by running our new cases with identical parameters to the dynamo simulation ``Case 4.00" from \citetalias{Matilsky2024}, which had $\sigma=76.4$ and $\pr=1.00$ and was thus firmly in the viscous spreading regime. A very well-confined statistically-steady tachocline then develops via the dynamo confinement scenario already described. We then restart the simulation from the established confined-tachocline state, but change the parameters to $\sigma=2.88$ and $\pr=0.250$, which allows radiative spreading to become significant. We consider both an MHD case, in which the dynamo from Case 4.00 keeps running, and an HD case, in which the magnetism is turned off. To allow each system to achieve a new thermally-equilibrated, statistically-steady state, the new HD case is run for 46.6 thermal diffusion times and the new MHD case is run for 3.11 magnetic diffusion times (corresponding to approximately 8 dynamo cycles, thereby allowing the skin effect to operate). In each case, we use impenetrable, stress-free, potential-field, and zero-entropy-gradient boundary conditions at both boundaries. The computational grid has the following resolution, where we denote the number of radial, colatitudinal, and longitudinal points by $N_r$, $N_\theta$, and $N_\phi$, respectively. For the viscously spreading MHD case (which sets the initial condition), $(N_r,N_\theta,N_\phi)=(192,384,768)$; for the radiatively spreading HD case, $(N_r,N_\theta,N_\phi)=(192,768,1536)$; for the radiatively spreading MHD case, $(N_r,N_\theta,N_\phi)=(192,512,1024)$. For full numerical details, see \citetalias{Matilsky2024}.

Figure \ref{fig:fields_and_flows} shows typical elements of the flow velocity $\vecu$ (via the vorticity, $\vecom\define\curl\vecu$)  and magnetic field $\vecb$ in these CZ--RZ dynamo systems. In the CZ, the axial ``convective" vorticity (i.e., $\omega_z^\prime\define\omega_z-\av{\omega_z}$, where $\av{\cdots}$ denotes an azimuthal average at a particular instant) traces the helical columnar convective rolls known as Busse columns (e.g., \citealt{Busse2002}). 
These columns drive large-scale shear (e.g., \citealt{Gilman1972, Matilsky2019}), ensure thermal-wind balance \citep{Matilsky2020b}, and produce the large-scale poloidal magnetic field via dynamo action (e.g., \citealt{Augustson2013}; \citetalias{Matilsky2022}).
The figure also shows the nonaxisymmetric poloidal magnetic field component $\btheta$ in the RZ, which is a result of the penetration of the dynamo's magnetic field due to the skin effect, and whose associated Maxwell stresses ultimately confine the tachocline.

\begin{figure}
	\centering
	\includegraphics[width=3.4375in]{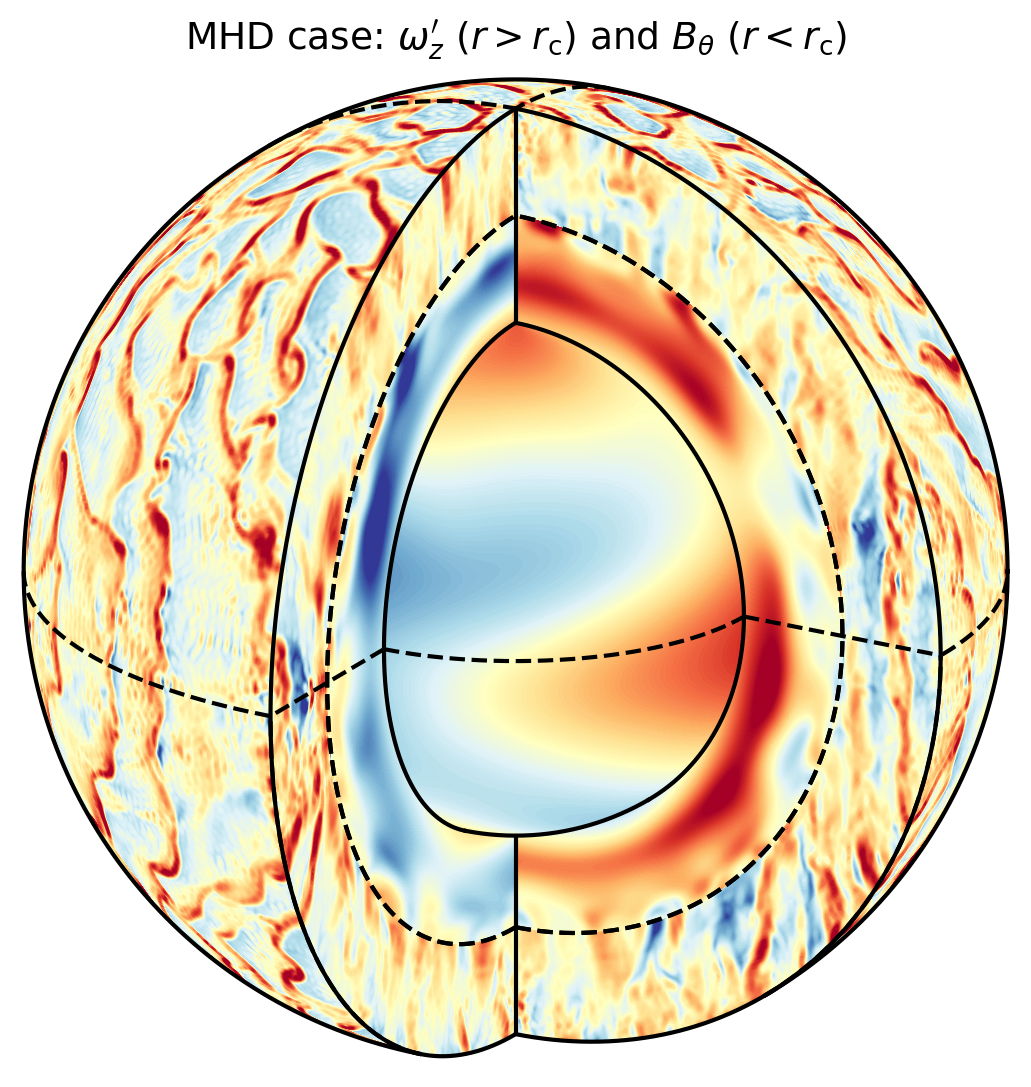}
	\caption{Three-dimensional spherical cutout showing a snapshot of the simulated flows ($\omzprime$ for $r>\rc$) and fields ($\btheta$ for $r<\rc$) in the MHD case. The spherical surfaces are close to the boundaries and the meridional surfaces are located $90^\circ$ apart. Each radial level is normalized separately by the rms of the appropriate quantity at that level. }
	\label{fig:fields_and_flows}
\end{figure}

We define the local rotation rate $\Omega$ via
\begin{equation}\label{eq:omega}
\Omega(r,\theta,t) \define \omzero + \frac{\av{\uphi}}{r\sin\theta},
\end{equation}
where $\phi$ is the azimuthal coordinate. Figure \ref{fig:tacho} shows the normalized differential rotation $\avt{\Omega}/\omzero -1$ in both the HD and MHD cases, where $\avt{\cdots}$ denotes a temporal average over the statistically-stationary and thermally-equilibrated state. The results are reminiscent of those in \citetalias{Matilsky2022}. In the MHD case, there is a confined tachocline, whereas in the HD case, the shear has spread throughout the whole RZ. Crucially, however, the spread of shear in the current simulations is radiative, as we now verify.

\begin{figure}
	\centering
	\includegraphics[width=3.4375in]{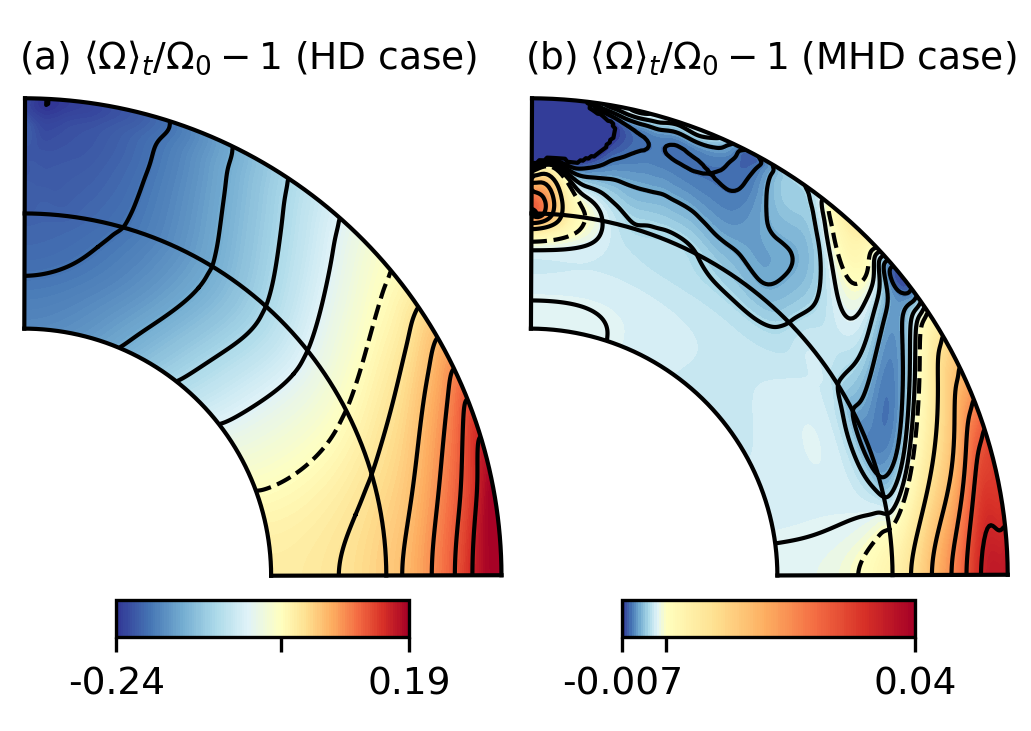}
	\caption{Simulated differential rotation $\avt{\Omega}/\omzero-1$, symmetrized about the equator, in (a) the HD case and (b) the MHD case. The colorbars are normalized separately for positive and negative values and the contour $\avt{\Omega}\equiv\Omega_0$ is dashed.}
	\label{fig:tacho}
\end{figure}

\begin{figure}
	\centering
	\includegraphics[width=3.4375in]{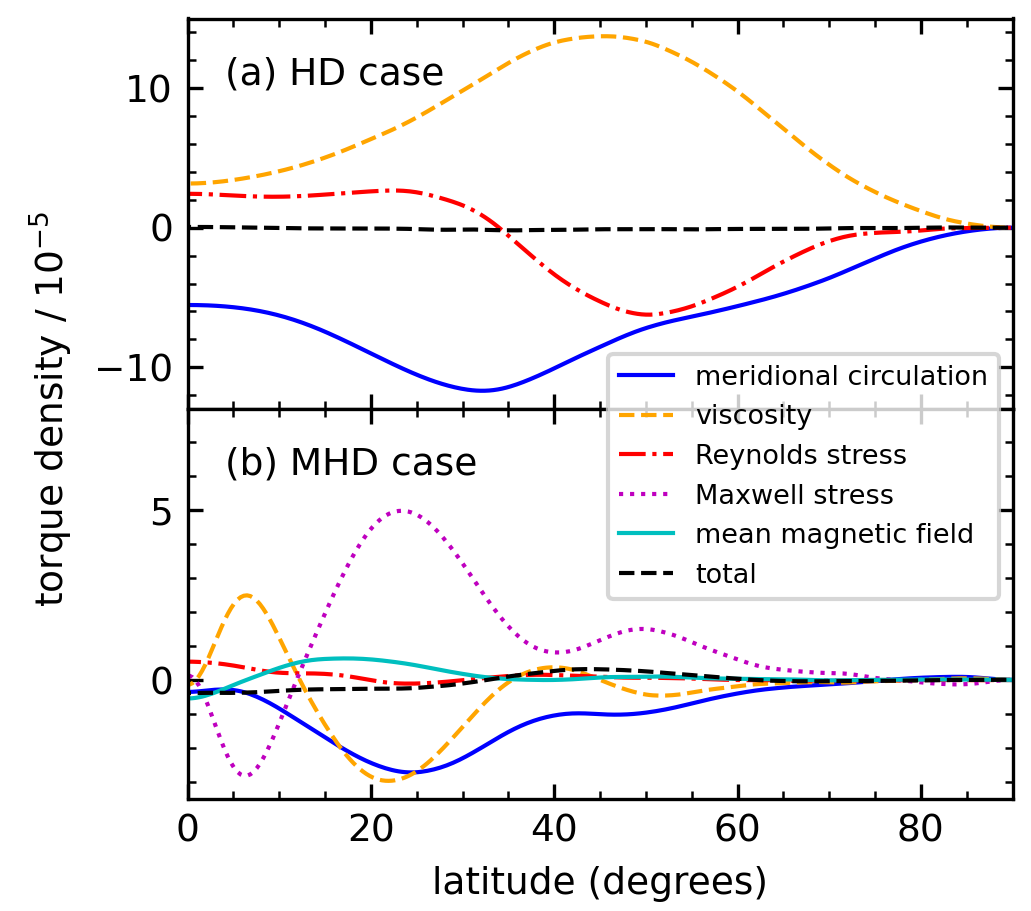}
	\caption{ Azimuthally and temporally averaged torque densities, symmetrized about the equator and radially averaged for $r/\rsun<0.696$ (i.e., below the approximate base of the overshoot layer), shown for (a) the HD case and (b) the MHD case. See \citetalias{Matilsky2024}'s Equation (17) for the full definitions of the torque densities marked in the legend.}
	\label{fig:torques}
\end{figure}

True radiative spreading consists of a specific set of terms being dominant in the five HD fluid equations as the shear burrows inward (see \citealt{Clark1973} or \citetalias{Spiegel1992}). This full balance largely appears to hold in our simulations, although we again must reserve a detailed exposition of this to our followup work. In this Letter, we instead follow the approach of \citet{Korre2024} and examine the steady-state torque balance. If the torque due to the burrowing meridional circulation ($\taumc$) is strong enough and is also attempting to imprint the CZ's differential rotation onto the RZ, then this indicates that radiative spreading is the primary process spreading shear into the RZ. Furthermore, radiative spread tends to spread the shear along cylinders in order to enforce the Taylor-Proudman constraint in the differential rotation ($\pderivline{\Omega}{z}\equiv0$). Radiative spreading of the tachocline is thus expected to be most powerful in the slowly rotating high-latitude regions, where the tachocline's isorotation contours are nearly orthogonal to the rotation axis.

Figure \ref{fig:torques} shows the steady-state torque balance as a function of latitude in the RZ for each case, after temporal, azimuthal, and radial averaging. Since the torques are balanced, $\taumc$ is the same magnitude as the viscous torque $\tauv$.  However, we note the key result that the torque due to the meridional circulation is negative ($\taumc<0$) at most latitudes in both cases. This suggests that radiative spreading is behaving exactly as expected, i.e., it is working to slow down the whole RZ in order to bring it into corotation (along cylinders) with the CZ above it. Crucially, as demonstrated by Figure \ref{fig:tacho}, the radiative spread has succeeded in slowing down the RZ at most latitudes in each case, with the net result that the whole RZ rotates significantly slower than its initial rotation rate from \citetalias{Matilsky2024}'s Case 4.00. The only exception appears to be the very low latitude regions of the HD case, in which the positive viscous torque ($\tauv>0$) appears to have spread the fast equator radially inward. 

Our overall interpretation of Figure \ref{fig:torques} is that, in the HD case, the radiative spread is largely unopposed and has imprinted the shear all throughout the RZ until finally being stopped by the viscosity. In the MHD case, by contrast, the Maxwell-stress torque $\taums$ associated with the nonaxisymmetric poloidal magnetic field (see Figure \ref{fig:fields_and_flows}) quickly stops the radiative spreading and confines a tachocline. 

It remains to confirm that the Maxwell stresses that produce the confining torque are a consequence of the same type of skin effect identified by \citetalias{Matilsky2024}. We demonstrate this by comparing the amplitude of the poloidal magnetic field $\bpol\define \brad\erad + \btheta\etheta$ (where the $\e$'s denote unit vectors) that is actually achieved in the RZ of the MHD case to the profile that the nonaxisymmetric skin effect predicts. The novel finding of \citetalias{Matilsky2024} was that for a nonaxisymmetric, quasicyclic $\bpol$, each component $\bpolmw$ at a given angular frequency $\omega$ and azimuthal wavenumber $m$ was exponentially damped in the RZ over its own characteristic skin-depth $\deltamw$. Because of the nonaxisymmetry, the frequency determining the skin-depth was Doppler-shifted by the the rigid rotation rate of the RZ, i.e.,
\begin{equation}\label{eq:deltamw}
		\deltamw\approx\sqrt{\frac{2\eta}{\omega-m\omrz}},
\end{equation}
where $\eta$ is the background magnetic diffusivity in the tachocline and $\omrz$ is the volume-average of $\Omega$ over the RZ.
Note that the details are slightly more complicated than this; see \citetalias{Matilsky2024}'s Equations 21--23. By summing the damping profiles over all frequencies, we can assess the agreement between the realized $|\bpol|^2$ and the skin-effect prediction. Figure \ref{fig:skin_effect} shows this comparison for the large-scale, mostly nonaxisymmetric $\bpol$. The agreement is exceedingly good for the large-scale field, which is strong evidence that the skin effect is responsible for the presence of the large-amplitude $\bpol$ in the RZ and the associated confinement of the tachocline.

\begin{figure}
	\centering
	\includegraphics[width=3.4375in]{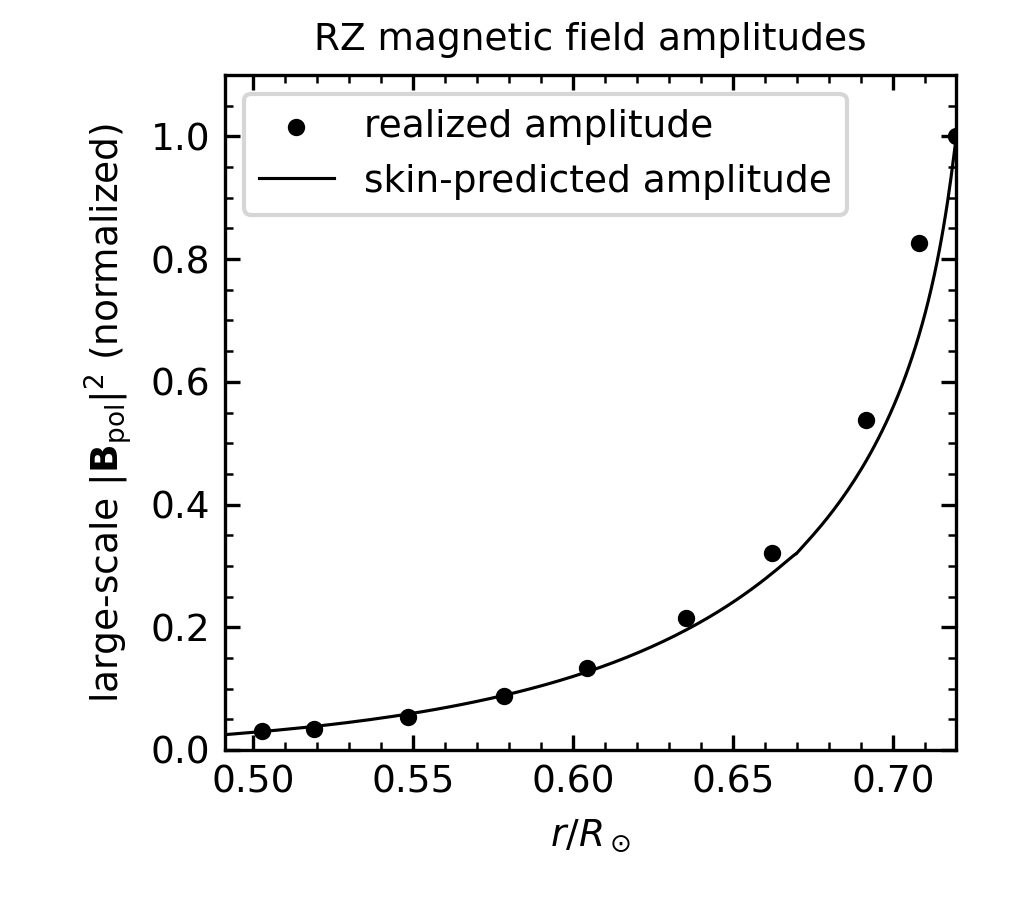}
	\caption{Dots: Realized amplitude of $|\bpol|^2$ in the RZ summed over the largest scales ($m=0,1,2$), where $m$ is the azimuthal wavenumber. Solid curve: Profile for $|\bpol|^2$ predicted by the nonaxisymmetric, quasicyclic skin effect. The skin-predicted amplitude comes from summing the exponential damping profiles $\approx$$e^{2(\rc-r)/\deltamw}$ over all $\omega$ and $m=0,1,2$, weighted by the value of $|\bpolmw|^2$ at the interface $r=\rc$. For full details, see \citetalias{Matilsky2024}'s Equations 21--23. }
	\label{fig:skin_effect}
\end{figure}

\section{Discussion}
The dynamo confinement scenario that we have demonstrated here, as with any theory, has its advantages and disadvantages. The essence of this scenario was first investigated by \citetalias{ForgcsDajka2001} (followed by \citealt{ForgcsDajka2002,ForgcsDajka2004,Barnabe2017}) in an idealized, two-dimensional, isolated RZ with imposed flows and fields at the top boundary. \citetalias{Matilsky2022} and \citetalias{Matilsky2024} showed that the core ideas of \citetalias{ForgcsDajka2001} were applicable in global nonlinear dynamo simulations of fully coupled CZ--RZ systems. Although the shear spread viscously instead of radiatively, these simulations self-consistently established a differentially-rotating CZ and a rigidly-rotating RZ separated by a statistically-stationary confined tachocline. Here, we have extended these ideas to demonstrate fully self-consistent dynamo confinement of a tachocline in the more solar-appropriate radiative spreading regime.  This is a major step forward due to the complexities of accessing this regime.

Another major advantage of this new dynamo confinement scenario is that the skin-effect principles at work here are far more flexible than originally conceived by \citetalias{ForgcsDajka2001}. The Doppler shifts caused by nonaxisymmetric dynamo modes moving relative to the RZ produce a variety of oscillation frequencies. The penetration of the associated confining Maxwell stresses could thus vary from very shallow to very deep. Indeed, for nonaxisymmetric fields almost corotating with the RZ (i.e., $\omega \sim m\omrz$), the penetration depth would be exceedingly deep according to Equation \eqref{eq:deltamw}. If this were to be the case in the Sun, the dynamo would have the potential to not only confine the tachocline, but also provide a pathway for the rigidification (and possibly even the spin-down) of the deep radiative interior. 

There are of course also reasons to be concerned with this dynamo confinement scenario. First, in all our simulations (including \citetalias{Matilsky2022} and \citetalias{Matilsky2024}), the latitudinal differential rotation in the CZ is substantially suppressed by the dynamo and therefore unrealistic in light of the strong differential rotation observed throughout the solar CZ. Nevertheless, we estimated in \citetalias{Matilsky2024} that a nonaxisymmetric $\bpol$ between a few G and a few hundred G in the tachocline could yield a Maxwell stress strong enough to prevent the radiative spread and yet not so strong that it disturbs the differential rotation in the CZ. Whether such a field strength is realistic could be assessed by future dynamo simulations at increasingly more solar-like parameters.

Second, it is somewhat troubling that the skin effect relies on significant magnetic diffusivity. Using microscopic values of $\eta$, \citetalias{Matilsky2024} estimated that a skin-depth larger than $\sim$$0.05$$\rsun$ (required for dynamo confinement) would require a Doppler-shifted cycle period of about 1.4 Gyr, which is an evolutionary timescale. Thus, for the skin effect to be significant, some sort of turbulent diffusivity might need to be invoked, even in the more flexible context of nonaxisymmetric, quasicyclic fields. Turbulent enhancement of the viscous and thermal diffusivities in the tachocline has long been assumed (e.g., \citealt{Zahn1992}) and is now reasonably well investigated in studies of stratified turbulence (e.g., \citealt{Cope2020,Chini2022,Shah2024,Garaud2025}). It therefore seems reasonable to assume that the magnetic diffusivity might be similarly enhanced (which, of course, was a central assumption in the original work by \citetalias{ForgcsDajka2001}). 

However, we must acknowledge that much is still unknown regarding the turbulent enhancement of any of the relevant diffusivities (viscous, thermal, and magnetic; for example, see the work by \citealt{Petrovay2003,Zhang2013,Jorgensen2018,Garaud2025}). Therefore, there is significant uncertainty in the understanding of turbulent radiative spreading, turbulent magnetic skin effects, and thus the nature of the true solar regime.  Nevertheless, we have now proven that the dynamo confinement scenario presented in this line of work can succeed in both of the original \citetalias{Spiegel1992} (non-turbulently-enhanced) viscous and radiative spreading regimes.



We therefore believe that, despite the stated concerns, the dynamo confinement scenario deserves further consideration, especially given the flexibility of the nonaxisymmetric skin effect. To this end, we have already performed a series of HD/MHD simulation pairs of CZ--RZ systems in the radiative spreading regime. These pairs, of which the HD/MHD pair presented in this Letter is one example, have various initial conditions,  background states, $\pr$, and $\sigma$. Using the larger simulation suite, we are investigating the dependence of the confinement mechanism and skin effect as the more solar-like (low $\sigma$, low $\pr$) regime is approached. The results of this broader study will be reported in a forthcoming, more comprehensive article. 

\begin{acknowledgments}
We thank P. Garaud, N. Featherstone, B. Hindman, C. Blume, J. Pedlosky, V. Skoutnev, and J. Toomre for helpful discussions. We thank an anonymous reviewer for highly insightful commentary. This work was primarily supported by the COFFIES DRIVE Science Center (NASA grant 80NSSC22M0162), with additional support from NSF award AST-2202253 and NASA grants 80NSSC18K1125, 80NSSC18K1127, 80NSSC19K0267,  80NSSC20K0193, 80NSSC21K0455, 80NSSC24K0270 and 80NSSC24K0125. Computational resources were provided by the NASA High-End Computing (HEC) Program through the NASA Advanced Supercomputing (NAS) Division at Ames Research Center. {\rayleigh} is supported by the Computational Infrastructure for Geodynamics (CIG) through NSF awards NSF-0949446 and NSF-1550901. For reproducibility, the {\rayleigh} input files and final checkpoints for each simulation are publicly accessible via Zenodo \citep{Matilsky2025b}.
\end{acknowledgments}


\end{document}